\documentclass[letterpaper, 10 pt, conference] {ieeeconf}  

\IEEEoverridecommandlockouts                              

\usepackage{graphicx}
\usepackage{cite}
\title{\LARGE \bf
A Classification Model Utilizing Facial Landmark Tracking to Determine Sentence Types for American Sign Language Recognition
}

\author{Janice Nguyen$^{1}$, Y. Curtis Wang$^{1}$
\thanks{$^{1}$ J. Nguyen is a graduating Master student at the Department of Electrical and Computer Engineering, California State University, Los Angeles and was advised by Y. Curtis Wang.}%
}

\begin{document}

\maketitle
\thispagestyle{empty}
\pagestyle{empty}

\begin{abstract}
   The deaf and hard of hearing community relies on American Sign Language (ASL) as their primary mode of communication, but communication with others who do not know ASL can be difficult, especially during emergencies where no interpreter is available. As an effort to alleviate this problem, research in computer vision based real time ASL interpreting models is ongoing. However, most of these models are hand shape (gesture) based and lack the integration of facial cues, which are crucial in ASL to convey tone and distinguish sentence types. Thus, the integration of facial cues in computer vision based ASL interpreting models has the potential to improve performance and reliability. In this paper, we introduce a simple, computationally efficient facial expression based classification model that can be used to improve ASL interpreting models. This model utilizes the relative angles of facial landmarks with principal component analysis and a Random Forest Classification tree model to classify frames taken from videos of ASL users signing a complete sentence. The model classifies the frames as statements or assertions. The model was able to achieve an accuracy of 86.5\%. 
\end{abstract}

\section{INTRODUCTION}
American Sign Language (ASL) is utilized by the deaf community as its primary mode of communication. However, deaf and hard of hearing individuals are often at risk for receiving inadequate treatment especially in healthcare due to communication barriers in a hearing centric society. Even though federal laws mandate hospitals  to provide interpreters, a majority of hospitals opt to use web-based video remote interpreting (VRI), which cause problems with communication between hearing physicians and hard of hearing patients due to issues such as poor video call quality, and lack of space in more crowded hospitals for the patient to sign and see the interpreter \cite{james_theyre_2022}. The poor quality of VRI services can be problematic, especially in emergencies where the patient needs immediate care. Thus, ongoing research in accurate, real time ASL interpreting models is vital.

ASL has its own unique linguistic structure and rules that dictate the meaning of individual signs and how to construct certain sentences. For instances, individual sign meaning and sentence formation is determined by five parameters: hand shape, palm orientation, facial expression, movement, and location relative to the body \cite{baker-shenk_american_1980}. A majority of current ASL interpreting models are essentially based on a signer's hand shape. These models perform ideally at isolating sign meanings, but this often leads to a limited set of signs that the model can interpret (\cite{vogler_asl_1998}, \cite{taskiran_real-time_2018}). Furthermore, these models perform poorly on more complex structures such as sentences and questions because the model is aimed towards interpreting individual signs instead of interpreting the meaning of entire sentences or questions. In cases where sentences have the same hand gestures but different meanings, the only way to differentiate these sentences is by facial cues. For example, "Are you hungry?" and "You are hungry" have the same hand gestures done in the same sequence and the only way to differentiate the two sentences is by raising eyebrows to indicate a yes or no question \cite{nguyen_facial_2012}. Thus, the lack of integrating facial expression recognition can cause poor performance in more complex structures.

Other ASL interpreting models based on facial expressions suggest there is potential in improving real time ASL interpreting integrating facial expressions, but the work in this field is limited due to the difficult task of tracking the dynamics of the subject’s constantly changing facial expressions and the potential loss of information from obstruction of the signer’s hands. Volger et al. addresses this problem by developing a 3D deformable model tracking system for the purpose of tracking faces of signers while they are signing in a video \cite{vogler_facial_2008}. However, this model requires time-consuming preprocessing of data before a model can be created. Another approach is to determine points on the subjects’ face and classify types of ASL signs based on that. Nguyen et al. developed a complex algorithm where 21 facial points of a subject’s face was determined and tracked by probabilistic principal component analysis and the results of the tracking were classified using a system based on Hidden Markov Models and another one based on a support vector machine \cite{nguyen_facial_2012}. While this model performs well, it is computationally expensive and time consuming to train and run due to its complex calculations. 

Our work uses the Random Forest model to simplify the computation, such that this classification can be performed on resource-constrained embedded systems or mobile devices. Munasinghe et al. \cite{munasinghe_facial_2018} also previously used a random forest model trained on facial landmark-based feature vectors data classified four facial emotions on normalized distances between points to account for factors that affect distances between the points, such as differing face sizes among subjects. 

In this paper, we propose a new approach to classify ASL sentences by transforming the facial landmarks using a relative angle-based approach. This new approach classifies signs as Assertions (AS) or Statements (ST)  \cite{baker-shenk_american_1980} while offering a simpler and computationally lean approach without sacrificing performance.  

\begin{figure*}
\begin{center}
\includegraphics[width=1\linewidth]{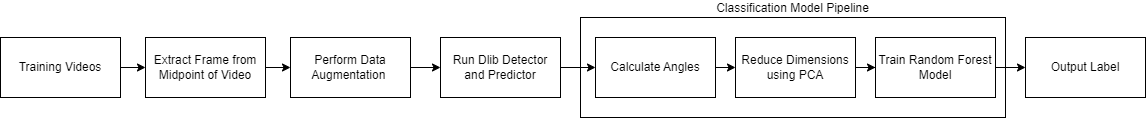}
\end{center}
   \caption{Training workflow for the proposed classification model pipeline.}
\label{fig:training_workflow}
\end{figure*}

\section{Proposed Model}

\begin{figure}
\begin{center}
\includegraphics[width=1\linewidth]{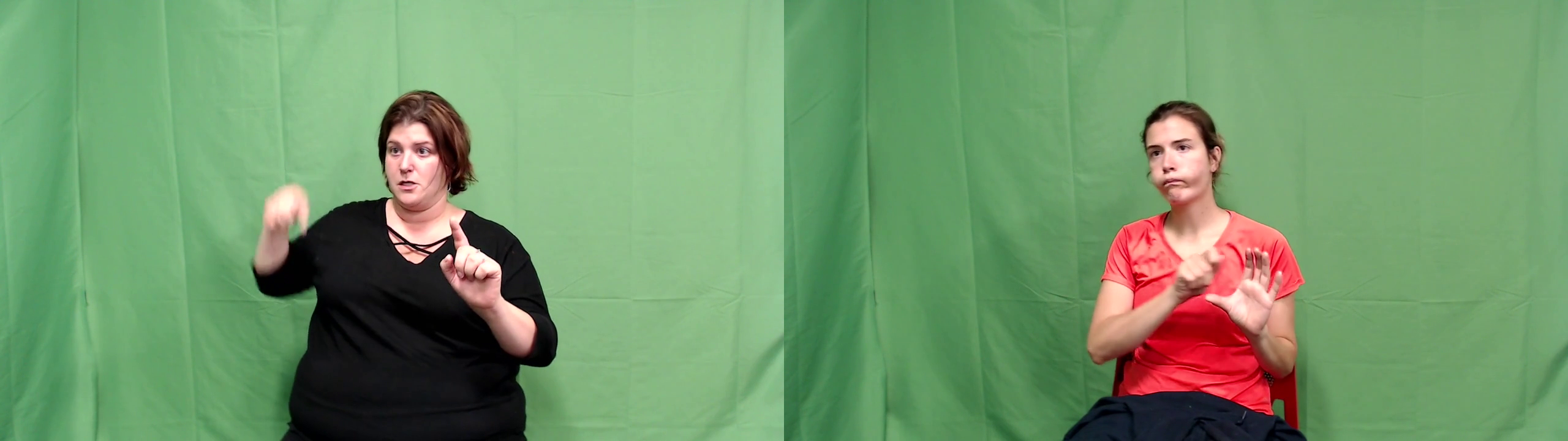}
\end{center}
   \caption{Left shows example of AS. Right shows example of ST. \cite{duarte_how2sign_2021}.}
\label{fig:aug_image}
\end{figure}

\subsection{Data Set}
\label{ssec:data}
For this project, we used the data set from How2Sign \cite{duarte_how2sign_2021} to train and test the model. We used a total of 173 videos of subjects signing sentences in ASL, where 121 videos were used for training and 52 videos were used for testing. The videos were classified into assertions (AS) and statements (ST). Assertions are considered sentences where the signer declared an action will occur while statements are sentences stating a fact. For every video, we extracted the midpoint frame and used that frame for training and testing the algorithm. We chose to use the midpoint frame because the grammatical markers that differentiate AS and ST sentences are usually found towards the latter part of a signed sentence \cite{baker-shenk_american_1980}. To increase the amount of training data used for this project, we applied data augmentation methods such as rotation, shifting, and resizing to the training data set. The degree of rotation, shifting and resizing were randomly chosen by the "ImageGenerator" function from the sklearn library. The data augmentation methods resulted in 3661 images and were all used for training. Testing was done on the remaining 52 videos.

\subsection{Extracting Facial Points using Dlib Library}
\label{ssec:dlib}
To calculate the facial points used for classification, we utilized the facial landmark detector and predictor from the Dlib library \cite{king_dlib-ml_2009} to detect the faces of the subjects and track key facial points. First, we need to detect and localize the subjects' faces before extracting facial points. To do so, we used the "get\_frontal\_face\_detector" function from the Dlib library which creates a facial detector that was developed using Histogram of Oriented Gradients (HOG) and Linear SVM classifier in conjunction with a image pyramid and sliding window method \cite{zhang_implementation_2020} to detect the subjects' faces before extracting facial points. 

After detecting the face, we used the "shape\_predictor" function based on the landmark estimator algorithm proposed by Vahid et al. \cite{kazemi_one_2014} and trained on the iBUG 300-W facial landmark dataset \cite{sagonas_300_2016} to determine the location of certain facial points on the subject's face. This function outputs 68 coordinates with the origin being at the top left of the image. 

\begin{figure}[t]
\begin{center}
\includegraphics[width=1\linewidth]{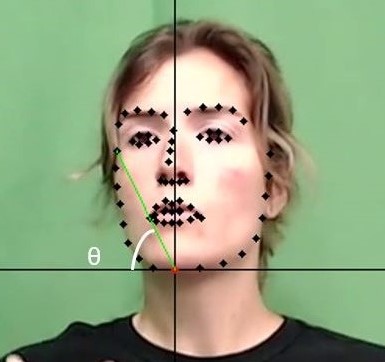}
\end{center}
   \caption{Example of Angle Calculation using the Red Point as the Origin on Image from How2Sign Dataset \cite{duarte_how2sign_2021}.}
\label{fig:angle_cal}
\end{figure}

\subsection{Angle Calculation}
After determining facial points, we calculate the feature vectors and angles using the output of "shape\_predictor." To begin with, the point at the center of the subject's chin (the red point in Figure \ref{fig:angle_cal}) is chosen to be the origin and all the other points is converted to the new origin's coordinate system before further calculations is done. Due to the inconsistencies in distances between the points from factors such as recording camera angle and differing subject's face sizes, we cannot consider the vectors alone and convert the vectors to angles using the following equation: 

\begin{equation}
    \theta_i = \arccos{\frac{x_0-x_i}{\sqrt{(x_0-x_i)^2+(y_0-y_i)^2}}}
\end{equation}

where $i$ is the $i^{th}$ point and $(x_0, y_0)$ is the red point on Figure \ref{fig:angle_cal}. Using angles ensures that factors such as inconsistencies in the subjects faces or rotation done by the subject during the video do not affect the model. Every image resulted in 67 angles that is later on reduced to 4 principal components. 

\subsection{Principal Component Analysis} 
To reduce the number of features used to train the model, we implement principal component analysis to reduce the 67 angles calculated to 4 principal components. Dimensionality reduction using PCA was done because it offers the potential to reduce complex data to the least amount of components while still retaining the significance of the dataset and has additional benefits such as using less computational power \cite{tipping_mixtures_nodate}. The PCA algorithm was implemented using the decomposition PCA function from the sklearn library \cite{pedregosa_scikit-learn_nodate}. 

\subsection{Random Forest Classification Algorithm}
After dimensionality reduction using PCA, we use the Random Forest Classification algorithm to classify each image. The Random forest model was chosen for its ability to reduce overfitting and handle large amounts of data. It works by creating an ensemble of decision trees and merging the results of each tree at the end \cite{biau_random_2016}. 

We implement the algorithm using the sklearn library from Python \cite{pedregosa_scikit-learn_nodate}. Pre-processing the data included running the Dlib detector and predictor to obtain 68 coordinates for every image and then angle calculation is done, resulting in 67 angles per image (we do not take into account  the angle between the origin and itself). We store the angles and classes for each image in .csv files. During training, we loaded the .csv files into the "pca" function from sklearn, resulting in 4 principal components. These 4 principal components were used to train a Random Forest Classification model and were inputted into the "RandomForestClassifier" function from sklearn. Testing was done on images that were not part of the training set.

\section{Results}
We trained and tested the model on principal components on videos from the How2Sign dataset \cite{duarte_how2sign_2021} and we trained the model to classify frames from videos of subjects signing complete sentences into 2 classes. Training was done on 70\% videos and data augmentation was used to increased the training data to 3661 images. Testing was done on the remaining 30\% of the videos that were not included in the training stage. Data augmentation was not done on the testing set. The model achieved an 86.5\% accuracy. Its average execution time over 10 trials was 4.353 ms per image. Running the Dlib process took 4.126 ms, the PCA calculation took 0.004 ms, and the random forest model took 0.223 ms per image. The entire process was run on Google Colab Pro+ using the Intel(R) Xeon(R) CPU @ 2.20GHz.  

\begin{table}
\centering
\begin{tabular}{ccccc}
\hline
\textbf{Class} & \textbf{TPR}   & \textbf{FPR}  & \textbf{TNR } & \textbf{FNR}         \\
\hline
\textbf{AS}        &  0.870  & 0.138 & 0.862 & 0.130     \\
\textbf{ST}        &  0.862  & 0.130 & 0.870 & 0.138     \\
\hline
 \end{tabular}
      \caption{Confusion Matrix.}
 \end{table}

 Previous works presented more complex, and computationally expensive means of classifying ASL sentence types using facial cues. These works often require a preliminary step where each frame regardless of whether the frame had the same signer were needed to be manually marked before undergoing further calculations used to train the model(\cite{nguyen_facial_2012}). Furthermore, due to the dynamic nature of the frames, other facial tracking models such as deformable models used for ASL interpreting require continuous complex mathematical analysis (\cite{vogler_analysis_nodate}) on the entire face of every frame of one video to be rotational invariant and to account for differences in facial size and shape. With our model, we introduce a simple, fast classification model that maintains rotational invariance through the usage of the facial landmarks from the Dlib library and angle calculation, and is computational efficient, with average execution time over 10 trials was 4.34 ms per image and a simple set up of just extracting the middle frame of each sentence. Looking at the confusion matrix, the TPR and TNR of AS and ST are balanced, indicating the model did not favor one class over the other.

\section{Conclusion}
Real time, autonomous ASL interpreting is a difficult task due to the complications in tracking and isolating facial expressions. This paper proposes a Random Forest Classification model trained on the principal components calculated when applying PCA on facial landmarks to classify frames from videos of subjects signing complete sentences in ASL. This model is designed to classify frames from videos of subjects signing complete sentences into assertions or statements. The "facial\_detector" and "shape\_predictor" functions from the Dlib library \cite{king_dlib-ml_2009} extracted coordinates of certain facial points. We calculated angles using these coordinates and used it to train the Random Forest Model. The model was able to correctly classify the testing data with a 86.5\% accuracy. Once the model is more robust, it can be trained to classify more types of sentences present in ASL syntax and be used in conjunction with previous hand gesture based models to implement real time ASL interpreting. 

\section{Compliance with ethical standards}
This research study was conducted retrospectively using human subject data made available in open access by How2Sign. Ethical approval was not required as confirmed by the license attached with the open access data.





\section*{ACKNOWLEDGMENT}
This work was supported by the CREST Center for Energy and Sustainability and the CREST Center for Advancement Towards Sustainable Urban Systems through National Science Foundation awards HRD-1547723 and HRD-2112554, respectively. 


\bibliographystyle{IEEEtran}
\bibliography{references.bib}

\end{document}